\documentclass[twoside,twocolumn,9pt]{article}
\usepackage{extsizes}
\usepackage[super,sort&compress,comma]{natbib} 
\usepackage[version=3]{mhchem}
\usepackage[left=1.5cm, right=1.5cm, top=1.785cm, bottom=2.0cm]{geometry}
\usepackage{balance}
\usepackage{mathptmx}
\usepackage{sectsty}
\usepackage{graphicx} 
\usepackage{lastpage}
\usepackage[format=plain,justification=justified,singlelinecheck=false,font={stretch=1.125,small,sf},labelfont=bf,labelsep=space]{caption}
\usepackage{float}
\usepackage{fancyhdr}
\usepackage{fnpos}
\usepackage[english]{babel}
\addto{\captionsenglish}{%
  
}
\usepackage{array}
\usepackage{droidsans}
\usepackage{charter}
\usepackage[T1]{fontenc}
\usepackage[usenames,dvipsnames]{xcolor}
\usepackage{setspace}
\usepackage[compact]{titlesec}
\usepackage{hyperref}

\usepackage{epstopdf}

\definecolor{cream}{RGB}{222,217,201}

\usepackage[sfdefault]{inter} 
\usepackage{hyperref}
\usepackage{latexsym}
\usepackage{mathdots}
\usepackage{siunitx}
\usepackage{longtable}
\usepackage{booktabs}
\makeatletter
\let\oldlt\longtable
\let\endoldlt\endlongtable
\def\longtable{\@ifnextchar[\longtable@i \longtable@ii}
\def\longtable@i[#1]{\begin{figure}[t]
\onecolumn
\begin{minipage}{0.5\textwidth}
\oldlt[#1]
}
\def\longtable@ii{\begin{figure}[t]
\onecolumn
\begin{minipage}{0.5\textwidth}
\oldlt
}
\def\endlongtable{\endoldlt
\end{minipage}
\twocolumn
\end{figure}}
\makeatother

\begin{document}

\pagestyle{fancy}
\thispagestyle{plain}
\fancypagestyle{plain}{
\renewcommand{\headrulewidth}{0pt}
}

\makeFNbottom
\makeatletter
\renewcommand\LARGE{\@setfontsize\LARGE{15pt}{17}}
\renewcommand\Large{\@setfontsize\Large{12pt}{14}}
\renewcommand\large{\@setfontsize\large{10pt}{12}}
\renewcommand\footnotesize{\@setfontsize\footnotesize{7pt}{10}}
\renewcommand\scriptsize{\@setfontsize\scriptsize{7pt}{7}}
\makeatother

\renewcommand{\thefootnote}{\fnsymbol{footnote}}
\renewcommand\footnoterule{\vspace*{1pt}%
\color{cream}\hrule width 3.5in height 0.4pt \color{black} \vspace*{5pt}} 
\setcounter{secnumdepth}{5}

\makeatletter 
\renewcommand\@biblabel[1]{#1}            
\renewcommand\@makefntext[1]%
{\noindent\makebox[0pt][r]{\@thefnmark\,}#1}
\makeatother 
\renewcommand{\figurename}{\small{Fig.}~}
\sectionfont{\sffamily\Large}
\subsectionfont{\normalsize}
\subsubsectionfont{\bf}
\setstretch{1.125} 
\setlength{\skip\footins}{0.8cm}
\setlength{\footnotesep}{0.25cm}
\setlength{\jot}{10pt}
\titlespacing*{\section}{0pt}{4pt}{4pt}
\titlespacing*{\subsection}{0pt}{15pt}{1pt}

\fancyfoot{}
\fancyfoot[LO,RE]{\vspace{-7.1pt}\includegraphics[height=9pt]{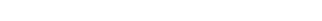}}
\fancyfoot[CO]{\vspace{-7.1pt}\hspace{13.2cm}\includegraphics{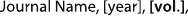}}
\fancyfoot[CE]{\vspace{-7.2pt}\hspace{-14.2cm}\includegraphics{head_foot/RF}}
\fancyfoot[RO]{\footnotesize{\sffamily{1--\pageref{LastPage} ~\textbar  \hspace{2pt}\thepage}}}
\fancyfoot[LE]{\footnotesize{\sffamily{\thepage~\textbar\hspace{3.45cm} 1--\pageref{LastPage}}}}
\fancyhead{}
\renewcommand{\headrulewidth}{0pt} 
\renewcommand{\footrulewidth}{0pt}
\setlength{\arrayrulewidth}{1pt}
\setlength{\columnsep}{6.5mm}
\setlength\bibsep{1pt}

\makeatletter 
\newlength{\figrulesep} 
\setlength{\figrulesep}{0.5\textfloatsep} 

\newcommand{\topfigrule}{\vspace*{-1pt}%
\noindent{\color{cream}\rule[-\figrulesep]{\columnwidth}{1.5pt}} }

\newcommand{\botfigrule}{\vspace*{-2pt}%
\noindent{\color{cream}\rule[\figrulesep]{\columnwidth}{1.5pt}} }

\newcommand{\dblfigrule}{\vspace*{-1pt}%
\noindent{\color{cream}\rule[-\figrulesep]{\textwidth}{1.5pt}} }

\makeatother

\twocolumn[
    \begin{@twocolumnfalse}
    \vspace{3cm}
    \sffamily
    \begin{tabular}{m{4.5cm} p{13.5cm} }
    
    \includegraphics{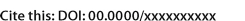} & \noindent\LARGE{\textbf{Magnetic
structure and properties of the honeycomb antiferromagnet
\ce{[Na(OH2)3]Mn(NCS)3}}} \\
    \vspace{0.3cm} & \vspace{0.3cm} \\
    
    & \noindent\large{Madeleine Geers,\textit{$^{a, b}$} Thomas B.
Gill,\textit{$^{c}$} Andrew D. Burnett,\textit{$^{d}$} Euan N.
Bassey,\textit{$^{e}$}$^{\ddag}$ Oscar Fabelo,\textit{$^{b}$} Laura
Cañadillas-Delgado,\textit{$^{b}$} Matthew J.
Cliffe,$^{\ast}$\textit{$^{a}$} $^{\dag}$} \\
    \includegraphics{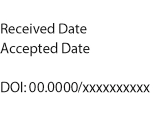} & \\
    \end{tabular}
    \end{@twocolumnfalse} \vspace{0.6cm}
]

\renewcommand*\rmdefault{bch}\normalfont\upshape
\rmfamily
\section*{}
\vspace{-1cm}


           \footnotetext{\textit{$^{a}$~School of Chemistry, University
Park, Nottingham, NG7 2RD, United
Kingdom~Email: matthew.cliffe@nottingham.ac.uk}}
               \footnotetext{\textit{$^{b}$~Institut Laue Langevin, 71
avenue des Martyrs CS 20156, 38042 Grenoble Cedex 9, France}}
               \footnotetext{\textit{$^{c}$~School of Electronic and
Electrical Engineering, University of Leeds, Leeds, LS2 9JT, UK}}
               \footnotetext{\textit{$^{d}$~School of Chemistry,
University of Leeds, Leeds, LS2 9JT, UK}}
               \footnotetext{\textit{$^{e}$~Yusuf Hamied Department of
Chemistry, University of Cambridge, Lensfield Road, Cambridge CB2 1EW}}
            \footnotetext{\ddag~ Present Address: Materials Department
and Materials Research Laboratory, University of California, Santa
Barbara, California 93106, United States of America}
    
\footnotetext{\dag~Electronic Supplementary Information (ESI) available: Additional
experimental details for synthesis, single crystal neutron diffraction
measurements and analysis, second harmonic generation, mCIF. Research
data and analysis notebooks are available at the Nottingham Research
Data Management Repository DOI:
10.17639/nott.7396.. See DOI: 10.1039/b000000x/}



\sffamily{\textbf{We report the magnetic structure and properties of a
thiocyanate-based honeycomb magnet \ce{[Na(OH2)3]Mn(NCS)3} which
crystallises in the unusual low-symmetry trigonal space group
\(P\overline{3}\). Magnetic measurements on powder samples show this
material is an antiferromagnet (ordering temperature
\(T_\mathrm{N,mag} = 18.1(6)\,\)K) and can be described by nearest
neighbour antiferromagnetic interactions \(J=-11.07(4)\,\)K. A method
for growing neutron-diffraction sized single crystals (\textgreater10
mm\(^3\)) is demonstrated. Low temperature neutron single crystal
diffraction shows that the compound adopts the collinear
antiferromagnetic structure with \(T_\mathrm{N,neut}= 18.94(7)\,\)K,
magnetic space group \(P \bar{3}'\). Low temperature second-harmonic
generation (SHG) measurements provide no evidence of breaking of the
centre of symmetry.}}\\


\rmfamily 

Honeycomb magnets, due to their high symmetry and two-dimensionality,
can host a wide range of magnetic phases, including monolayer
ferromagnetism in \ce{CrI3}\citep{Huang2017}, proximate Kitaev spin
liquid states in \ce{RuCl3} and \ce{Na2Co2TeO6}
\citep{sears_magnetic_2015, yaoExcitationsOrderedParamagnetic2022},
complex spin textures in \ce{FeCl3}\citep{gao_spiral_2022}, and
magnetoelectricity in
\ce{Co2Mo3O8}.\citep{reschkeConfirmingTrilinearForm2022}

Molecular magnets are of particular interest for low-dimensional
magnetism as it is often more straightforward to produce well-isolated
low-dimensional connectivity, and the molecular ligands offer the
potential for a high degree of
tunability.\citep{thorarinsdottirMetalOrganicFramework2020} Honeycomb
molecular magnets are no exception to this, particularly those based on
chelating ligands such as
(\ce{ox^{2-}}=\ce{C2O4^{2-}}),\citep{zhongFerromagneticHeteroMetal1990, pellauxMolecularBasedMagnetismBimetallic1997, mathoniereMolecularbasedMixedValency1994}
tetraoxolene
(\ce{C6O4R2^{$n$-}})\citep{jeon2DSemiquinoneRadicalContaining2015}
amongst
others.\citep{rodriguez-dieguezSelfAssembledCationicHeterochiral2007, curelyThermodynamicsTwodimensionalHeisenberg1998}
These structures often have multiple different paramagnetic species
producing ferrimagnetic order, whether this is alternating metal sites
\ce{A[M^{II}M^{III}](ox)3},\citep{zhongFerromagneticHeteroMetal1990}
\ce{A[M^{II}M^{III}](C6O4R2)3}\citep{abherveOneDimensionalTwoDimensionalAnilateBased2014, atzoriFamilyLayeredChiral2013}
or the presence of radical ligands in
\ce{A_2[M2(C6O4R2)3]},\citep{jeon2DSemiquinoneRadicalContaining2015}
where A is charge balancing cation (typically alkylammonium). Equally,
distorted honeycomb structures, with multiple different exchange
pathways are
common.\citep{curelyThermodynamicsTwodimensionalHeisenberg1998, wangSolventTunedAzidoBridgedCo22006, goherSynthesisStructuralCharacterisation2000a}
Examples of ideal honeycomb lattice magnets are thus comparatively rare.

\begin{figure}
\hypertarget{fig:intro}{%
\centering
\includegraphics{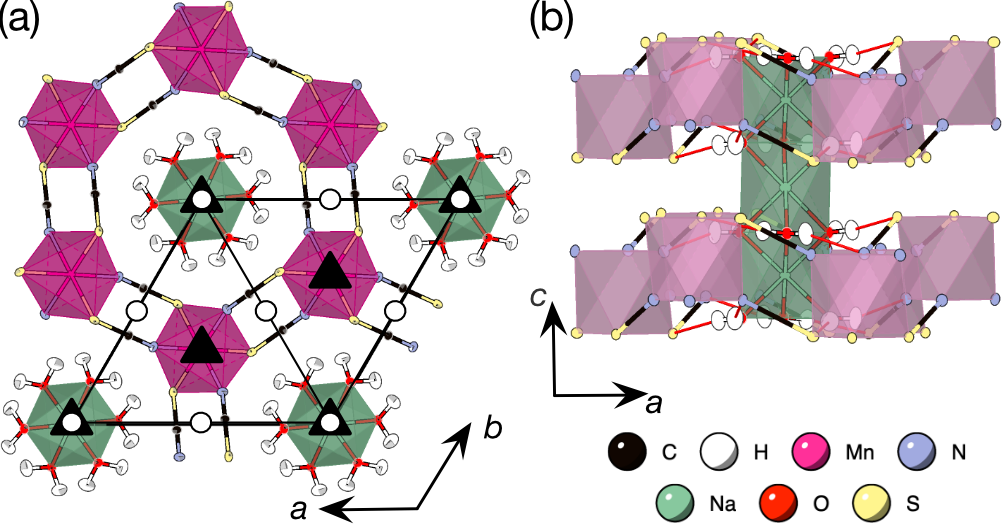}
\caption{Neutron single crystal structure of \ce{[Na(OH2)3]Mn(NCS)3}
(D19 at the ILL, 25 K).\citep{geersILLExperiment5411167} (a) View along
the \(c\) axis. Space group diagram for \(P \bar{3}\) overlaid,
\(\overline{1}\): white circle, \(3\): black triangle. (b) View along
\(b\) axis. Hydrogen bonds shown as red lines.}\label{fig:intro}
}
\end{figure}

One promising material family to search for honeycomb magnets are the
metal thiocyanates, which show a range of low dimensional magnetic
properties due to the layered structural connectivity and have
interactions as strong as comparable
halides.\citep{Bassey2020, geersNoncollinearMagnetismPostperovskite2023}
In this paper we investigate the magnetic properties of
\ce{[Na(OH2)3]Mn(NCS)3} (\ce{NaMn(NCS)3.3H2O}),\citep{Biedermann1998}
formed of ideal two-dimensional honeycomb \ce{[Mn(NCS)3]-} layers with
the hexagonal void filled by 1D \ce{[Na(OH2)3]+} rods
{[}Fig.~\ref{fig:intro}{]}. It crystallises in the low-symmetry but
still trigonal \(P\overline{3}\) space-group, comparatively rare in
inorganic materials (fewer than 0.25\% of ICSD
structures\citep{zagoracRecentDevelopmentsInorganic2019}). As a result,
the Mn(II) sites do not lie on inversion centres, which raises the
possibility that magnetic ordering may break the centre of symmetry and
hence induce simultaneous magnetic and electric order (Type II
multiferroicity).\citep{khomskii_classifying_2009} Although the
structure of this compound has been reported, because it is very
sensitive to humidity very little is known about its physical
properties.\citep{Biedermann1998} We report a method for the synthesis
of large \ce{[Na(OH2)3]Mn(NCS)3} single crystals (12 mm\(^3\)) of
suitable quality for neutron diffraction measurements, which mitigates
the strong humidity dependence of this compound. Using a combination of
bulk magnetic property measurements, low-temperature single crystal
neutron diffraction and SHG measurements, we have determined its
magnetic properties. We find that \ce{[Na(OH2)3]Mn(NCS)3} orders with
the classical honeycomb antiferromagnetic ground state, ordering into
the centrosymmetric collinear antiferromagnetic \(P\bar{3}'\) space
group.

We synthesised \ce{[Na(OH2)3]Mn(NCS)3} using a method adapted from that
of \citet{Biedermann1998}, using the salt metathesis of manganese(II)
sulfate and barium thiocyanate with additional sodium thiocyanate in
aqueous solution in a \(1:1:1\) ratio. After filtering the precipitated
\ce{BaSO4} and removing the water, a pale green-yellow powder was
obtained, deliquescent in ambient conditions, which can be recovered
from its own solution by heating at no more than \(60\)°C. The powder
sample is also sensitive to dry environments, and will decompose to
\ce{Mn(NCS)2} and NaNCS on heating.

On concentration \emph{in vacuo}, aqueous solutions of
\ce{[Na(OH2)3]Mn(NCS)3} change colour sequentially from colourless to
pale pink, pale green and eventually to aqua blue {[}ESI Fig. 1{]}. We
found that the optimal concentration for crystal growth is when the
solution is an intense green colour, just before it begins turning blue.
More concentrated, blue, solutions very rapidly crystallised, which led
to many smaller crystals. Less concentrated solutions were not
sufficiently supersaturated for crystal growth {[}ESI Fig. 2{]}. On
leaving this green solution to stand for 24 h at \(7\)°C we obtained
large regular-hexagonal crystals. As the supersaturated solution absorbs
water from the air, the concentration falls below the level required for
optimal crystal growth after 24 h. The harvested crystals from this
initial crystallisation were then used as seeds for future
crystallisation with the growth solution being reconcentrated to the
optimal concentrations. After nine iterations of seeded crystallisation
we obtained pale green, hexagonal crystals of \ce{[Na(OH2)3]Mn(NCS)3}
suitable for neutron diffraction experiments: \(4\times3\times1\)
mm\(^3\) {[}Fig.~\ref{fig:mag_struct} (\(\mathrm{a}\)){]}. Due to the
humidity sensitivity of this compound, we explored a variety of
conditions for storage using saturated aqueous salt solutions to control
the atmosphere: \(54\)\% (\ce{Mg(NO3)2}\(_{(\mathrm{aq.})}\)), \(39\)\%
(\ce{NaI}\(_{(\mathrm{aq.})}\)), \(32\)\%
(\ce{CaCl2}\(_{(\mathrm{aq.})}\)) and \(23\)\%
(\ce{KCH3COO}\(_{(\mathrm{aq.})}\)).\citep{greenspan_humidity_1977, obrien_control_1948}
We found that the crystals were indefinitely stable at \(32\)\% humidity
(\textgreater23 months). We confirmed the phase purity of powder samples
with powder X-ray diffraction and the quality of smaller single crystals
was checked using ambient temperature X-ray single crystal diffraction.

\begin{figure}
\hypertarget{fig:MPMS}{%
\centering
\includegraphics{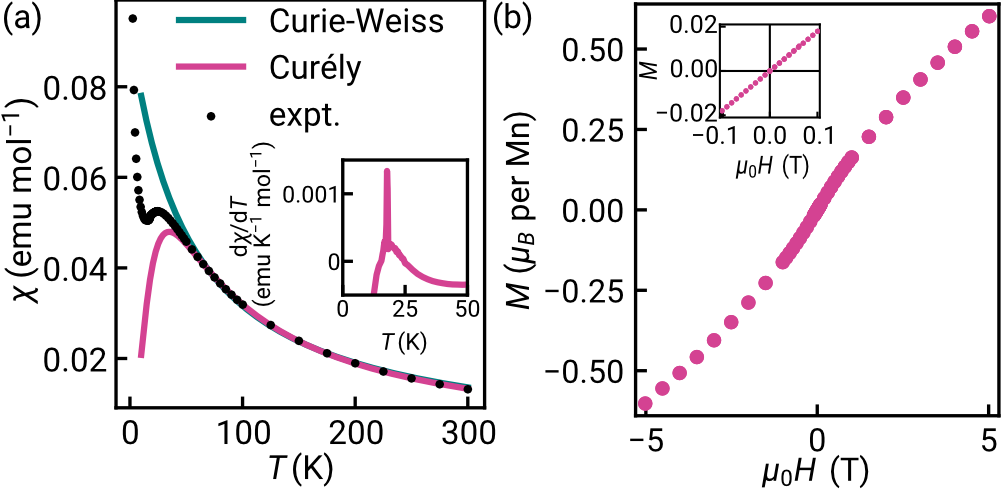}
\caption{(a) Magnetic susceptibility, \(\chi\) measured from 2-300 K
(black), with the high temperature fits using the Curie-Weiss (teal) and
Curély equation (pink) for a classical honeycomb magnet. Inset:
temperature derivative of the susceptibility in the vicinity of the
transition. (b) Isothermal magnetisation data, measured between
\(+5.00\) and \(-5.00\) T at \(2\) K. Inset: close-up of data between
\(+0.1\) and \(-0.1\) T}\label{fig:MPMS}
}
\end{figure}

Previous work has established that manganese thiocyanate compounds show
moderately strong magnetic interactions
\citep{Bassey2020, geersNoncollinearMagnetismPostperovskite2023}, and so
to investigate the magnetic properties of \ce{[Na(OH2)3]Mn(NCS)3} and
uncover any magnetic phase transitions, we carried out magnetic
susceptibility (\(\mu_0H = 0.01\) T) and isothermal magnetisation
measurements on a powder sample (\(2\) K, \(-5\) to \(+5\) T). We found
that the magnetic susceptibility increases on cooling until a broad
maximum is reached around \(25\) K. Below this temperature, the
susceptibility decreases until a sharp drop at the ordering temperature,
\(T_\mathrm{N} = 18.1(6)\) K {[}Fig.~\ref{fig:MPMS}\((\mathrm{a})\){]}.
The temperature dependence of the moment was fitted to the Curély model
of a nearest neighbour classical honeycomb
magnet\citep{curelyThermodynamicsTwodimensionalHeisenberg1998}, using
the following Hamiltonian:

\begin{equation}\label{eqn:ham} 
\mathcal{H} = -\frac{1}{2} \sum_{i,j} J S_i \cdot S_j,
\end{equation}

where \(|S|=\sqrt{S(S+1)}\), and the sum is only over nearest
neighbours. In this convention, antiferromagnetic superexchange will
give negative \(J\). We fitted the data over the full temperature range,
with an additional \(1/T\) term to account for paramagnetic impurities
due to sample hydration, giving \(J=-10.0(2)\,\)K and \(g=1.8\) with
7.0(2)\% paramagnetic impurities. We also fitted just the high
temperature regime (\(T>60\,\)K), using both the Curély model, giving
\(J=-11.07(4)\,\)K, \(g=2.02\), and using the Curie-Weiss model which
gives a Curie-Weiss temperature, \(\theta_\mathrm{CW} = -51.4(1.4)\,\)K,
equivalent to \(J=-17.4(5)\,\)K and \(g=2.09\), the overestimated
Curie-Weiss temperature being typical of analysis of low-dimensional
magnets. The presence of a small fraction of hydrated impurity means the
extracted values of \(g\) are unreliable.

The isothermal magnetisation does not saturate up to \(5.00(1)\) T,
consistent with the moderately strong antiferromagnetism, and there is
no observable hysteresis within the error limits of the measurements
(\textless1mT) {[}Fig.~\ref{fig:MPMS}(b){]}. A change in the gradient at
\(2.0(3)\) T is suggestive of a spin reorientation transition {[}ESI
Fig. 8{]}.

The combination of the negative \(\theta_\mathrm{CW}\) and lack of
hysteresis in the isothermal magnetisation data, suggests that
\ce{[Na(OH2)3]Mn(NCS)3} has ground-state antiferromagnetic order. If
there is a ferromagnetic component due to spin canting, which could
explain the up-turn in the low temperature susceptibility data, it must
be small in magnitude.

\begin{figure}
\hypertarget{fig:mag_struct}{%
\centering
\includegraphics{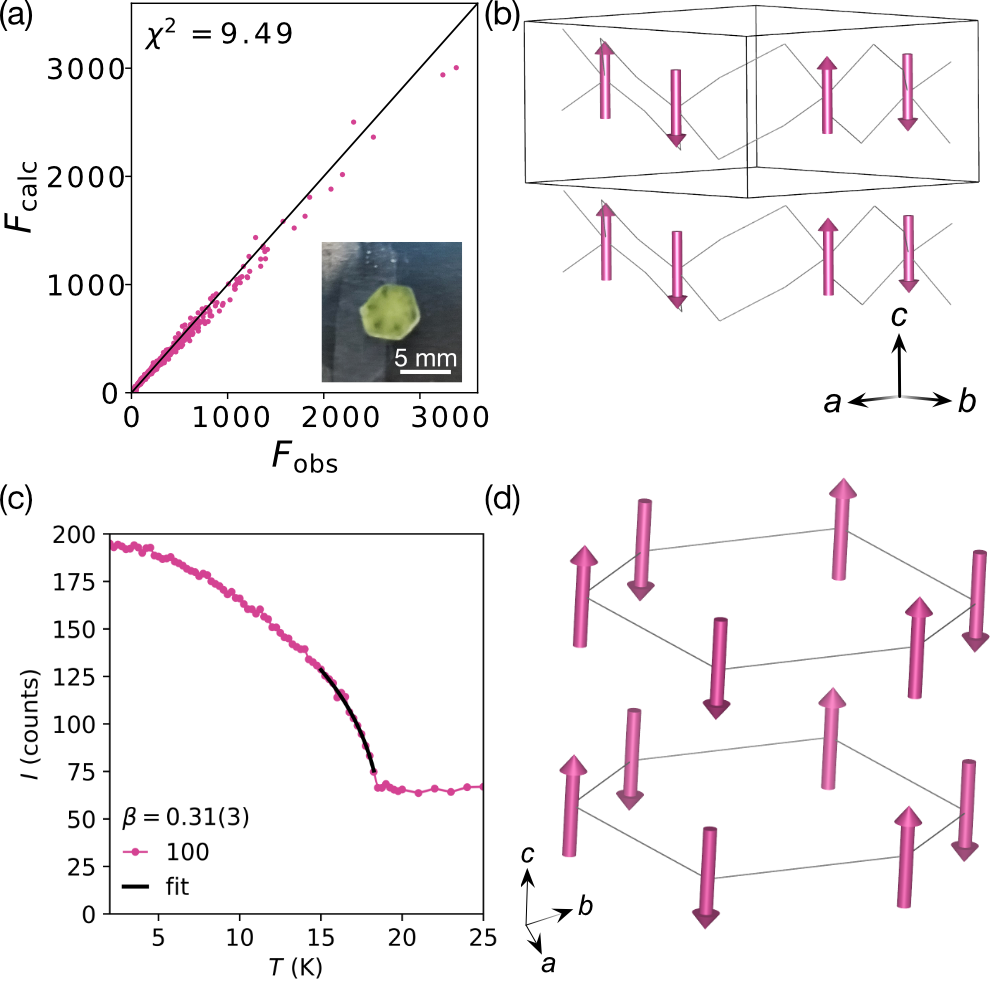}
\caption{Summary of the refined magnetic structure in \(P \bar{3}'\) at
\(2\) K (D19, ILL).\citep{geersILLExperiment5411167} (a)
\(F_{\mathrm{obs}}\) against \(F_{\mathrm{calc}}\) plot. Inset: the
measured single crystal. (b) Magnetic structure viewed along the
{[}110{]} direction showing the antiferromagnetic arrangement of moments
(pink arrows). {[}Mn(NCS)\(_3\){]}\(^-\) framework is shown as a
wireframe. (c) Temperature dependence of the \(100\) Bragg reflection,
which has a significant magnetic contribution, fitted to a power law
(black line), \(\beta=0.31(3)\). (d) Alternate view of the magnetic
structure, with the magnetic lattice (connecting Mn atoms)
shown.}\label{fig:mag_struct}
}
\end{figure}

To determine the low temperature nuclear and magnetic structure of
\ce{[Na(OH2)3]Mn(NCS)3}, single crystal neutron diffraction measurements
were performed (D19, ILL).\citep{geersILLExperiment5411167} Full
datasets were collected above (\(25\) K) and below (\(2\) K) the
ordering temperature of \(T_\mathrm{N} = 18.1(6)\) K. The data at \(25\)
K (\(\lambda = 1.45\) Å, \(631\) independent reflections) could be
integrated with the same space group as the reported structure at
ambient temperature, \(P\bar{3}\), indicating the absence of any
structural transitions. The atom positions and anisotropic displacement
parameters were refined freely, with a final \(\chi^2 = 23.20\),
obtained by refinement using the FullProf
programme.\citep{rodriguez-carvajal_recent_1993} From each \ce{H2O}
molecule one hydrogen is directed towards a nitrogen, whilst the other
hydrgoen is positioned towards a sulfur atom with \(\angle\)
\ce{O-H}\(\cdots\)N \(= 153.291(2) ^{\circ}\) and \(\angle\)
\ce{O-H}\(\cdots\)S \(= 164.258(2) ^{\circ}\). The distances of the
hydrogens to the acceptor atoms are
\emph{d}\(_\mathrm{H1 \cdots N} = 2.355(5)\) Å and
\emph{d}\(_\mathrm{H2 \cdots S} = 2.326(7)\) Å, consistent with previous
reports.\citep{Lommerse1998}

In the low temperature dataset (\(2\) K) all the Bragg reflections
(\(651\) unique reflections) could be indexed by the nuclear space group
\(P \bar{3}\), implying the propagation vector is \textbf{k} =
\((0, 0, 0)\). The potential magnetic space groups with this propagation
vector were identified using the Bilbao Crystallographic
Server,\citep{aroyo_crystallography_2011, gallego_magnetic_2012, perez-mato_symmetry-based_2015}
\(P\bar{3}'\), \(P\bar{3}\), \(P3\), \(P\bar{1}'\), \(P\bar{1}\) and
\(P1\) in BNS notation.\citep{gallego_magnetic_2012} We initially
focussed on the three maximal symmetry magnetic space groups:
\(P\bar{3}'\), \(P\bar{3}\) and \(P3\). Of these, the \(P\bar{3}\)
magnetic space group only allows for a collinear ferromagnetic order and
so was neglected as it is inconsistent with our magnetic property
measurements; \(P\bar{3}'\) permits only collinear antiferromagnetic
order and \(P3\) allows antiferromagnetic, ferromagnetic or
ferrimagnetic arrangement of the moments as it contains two distinct
Mn(II) sites.

Refinements were carried out with both the \(P\bar{3}'\) and \(P3\)
space groups. The results of the refinements provided comparable models
with similar refinement statistics, \(\chi^2 = 9.49\) (\(P\bar{3}'\))
and \(\chi^2 = 9.8\) (\(P3\)) {[}ESI Figs. 6 \& 7{]}. The significant
difference between these models is the loss of the inversion centres in
the \(P3\) space group, which produces the two unique Mn sites. In both
structures the Mn atoms are positioned on a 3-fold rotation axis and
thus the moments are constrained to lie along \emph{c}. The
antiferromagnetic \(P\bar{3}'\) model has a moment size, \(gS\), of
\(4.9(2) \; \mu_{\mathrm{B}}\) per Mn, whereas in the \(P3\) model, the
magnitude of the two moments are \(-3.4(3)\) and
\(5.4(2) \; \mu_{\mathrm{B}}\). These moments were closely correlated,
and were only stable during the refinements when the moments were
constrained to refine between \(-5.0\) and \(5.0 \; \mu_{\mathrm{B}}\).
This resulted in moments of \(-4.0(2)\) and
\(3.9(2) \; \mu_{\mathrm{B}}\). Within error these moments have the same
magnitude, although any uncompensated moment would result in a net
magnetisation along the \emph{c} axis. We also investigated the
potential for lower symmetry \(\textbf{k}=0\) orderings, with canted
moments or significant magnetostructural distortions, which would
require triclinic symmetry. We integrated the data with triclinic
symmetry, which produced a cell which was trigonal within error, and
then refined the magnetic and nuclear structures in \(P \bar{1}'\)
{[}ESI Fig. 4; ESI Table 1{]}. This model has a single Mn site,
\(M = 4.1(2) \; \mu_{\mathrm{B}}\), with the moment pointing primarily
along the \emph{c} axis with a significant canting, though the errors in
this model prevent reliable determination of moment direction. The
atomic coordinates and displacement parameters were refined freely,
however there was no evidence beyond error of any symmetry lowering
beyond trigonal symmetry and the refinement fit was significantly worse
than the for the higher symmetry models, with \(\chi^2 = 51.3\). We
therefore concluded that the triclinic models were not required to
describe this system.

To gain more detailed understanding of the transition we collected data
between \(2\) and \(25\) K with \(0.25\) K steps to follow specific
reflections. We focussed in particular on the \(100\) Bragg reflection,
which has significant magnetic contribution in the magnetically ordered
stated. We fitted the intensity of this reflection to a power law:

\begin{equation}\label{eqn:tn_powerlaw} 
I = M^2 = A(T_\mathrm{N} - T)^{2\beta}+C,
\end{equation}

\noindent where \emph{A} is a proportionality constant, \(T_\mathrm{N}\)
is the ordering temperature, \(\beta\) is a critical exponent and \(C\)
the nuclear scattering intensity. We found for the \(100\) reflection
that \(\beta = 0.31(3)\) and \(T_\mathrm{N} = 18.94(7)\) K
{[}Fig.~\ref{fig:mag_struct}(c){]}, which lies between the theoretical
results for a two-dimensional Ising antiferromagnet (\(\beta=0.125\))
and a three-dimensional Heisenberg antiferromagnet
(\(\beta=0.367\)).\citep{blundell_magnetism_2001} This suggests that
there is some degree of low dimensional character even in the vicinity
of the transition, as found for other two dimensional magnets:
\ce{FeBr3} (\(\beta = 0.324\)),\citep{cole_extreme_2023} \ce{NiCl2}
(\(\beta = 0.27\))\citep{lindgard_spin-wave_1975} and \ce{FeCl2}
(\(\beta = 0.29\)).\citep{yelon_magnetic_1972} Other Bragg reflections,
for example the \(235\) reflection, show almost no change in intensity
with temperature as they have very limited contribution from magnetic
scattering {[}ESI Fig. 9{]}. The nuclear structure at \(2\) K is very
similar to that at \(25\) K, with no statistically significant
deviations in bond lengths.

Our refinements suggested that the magnetic structure is most likely the
centrosymmetric antiferromagnetic \(P\bar{3}'\), rather than the
noncentrosymmetric weak ferromagnetic \(P3\). To confirm this model and
to determine if \ce{[Na(OH2)3]Mn(NCS)3} is centrosymmetric in its ground
state, we therefore carried out second harmonic generation measurements
at 4 K and room temperature, along with variable temperature
measurements between 4 and 30 K. Optical excitation was performed using
Spectra Physics Spitfire Ace PA amplified laser system (40 fs pulses at
a 1 KHz repetition rate) at 800 nm (\textless{} 1 W average power) and
400 nm (\textless{} 500mW). We found no evidence of SHG signal
generation, suggesting the compound is centrosymmetric. The results
suggest that \ce{[Na(OH2)3]Mn(NCS)3} retains its inversion centres,
therefore ordering in the \(P\bar{3}'\) magnetic space group, and
providing further evidence that this compound is a good example of a
classical Heisenberg honeycomb antiferromagnet.

The \ce{[Mn(NCS)3]-} layers of \ce{[Na(OH2)3]Mn(NCS)3} stack uniformly
along the \emph{c} axis, in contrast to the ABC stacking found in
\ce{NaMnCl3}.\citep{van_loon_crystal_1973} We found no evidence of
stacking faults, slip-stacking transitions or symmetry-lowering due to
stacking, in contrast to van der Waals materials such as \ce{MnPS3} and
and \ce{RuCl3} which have monoclinic symmetry because of the layer
sequence,\citep{ressoucheMagnetoelectricTextMnPS2010, johnson_monoclinic_2015}
This is likely due to the infinite rod \ce{[Na(OH2)3]^+_$n$}
countercation, which forces hydrogen bonds to the framework and force
the layers to stack in alignment.

The importance of the trigonal symmetry of the cation, together with its
templating hydrogen bonds, can be seen by comparison with
{[}1,3-Im{]}Mn(NCS)\(_3\) (1,3-Im = 1-ethyl-3-methyl imidazolium) which
has the bulky 1,3-Im cation as the charge balance to the
\ce{[Mn(NCS)3]-} honeycomb.\citep{Peppel2019} The 1,3-Im cation has no
strong hydrogen bond donors and its size and bulk means the framework
forms irregular hexagons. This distortion away from perfect symmetry is
often found in molecular framework honeycomb magnets, for example
\ce{Co(N3)2(bpg). DMSO} (bpg = 1,2-dipyridine-4-ethane-1,2-diol, DMSO =
\ce{SO(CH3)2}).
\citep{goherSynthesisStructuralCharacterisation2000a, wangSolventTunedAzidoBridgedCo22006}
The \ce{[Na(OH2)_3]+} countercation rod is not unique to
\ce{[Na(OH2)3]Mn(NCS)3}, occurring in a handful of other compounds,
\emph{e.g.}
\ce{[Na(OH_2)_3]_2[TeBr6]}.\citep{abrielKristallstrukturNaH2O1983}
Neutron diffraction allows a close look at the H-bond interactions
between this cation and the framework.

The magnetic properties of \ce{[Na(OH2)3]Mn(NCS)3} are quite distinct
from the halide analogue \ce{NaMnCl3}, which orders at
\(T_\mathrm{N} = 6.5\) K with ferromagnetic layers, coupled
antiferromagnetically.\citep{fedoseeva_magnetic_1980}
\ce{[Na(OH2)3]Mn(NCS)3} has an ordering temperature almost three times
as high as \ce{NaMnCl3} (\(T_\mathrm{N} = 18.1(6)\) K) and the net
magnetic interactions are also stronger,
\(|\theta_\mathrm{CW}| = -51(1)\) K, compared to \ce{NaMnCl3},
\(|\theta_\mathrm{CW}| = -4.2\) K. The increase strength of interactions
in the thiocyanate compared to the halide despite the significantly
longer superexchange pathway (\ce{Mn-Cl-Mn} compared to \ce{Mn-NCS-Mn})
is consistent with investigations of the binary
compounds.\citep{Bassey2020} This difference may be due to the harder
\ce{NCS-} ligand having better orbital overlap with \ce{Mn^{2+}} and the
single-orbital nature of the superexchange pathway, rather than the
two-orbital pathway in manganous chlorides.

Our results establish that \ce{[Na(OH2)3]Mn(NCS)3} adopts the
\(P \bar{3}'\) space group, but it is worth exploring implications of
the \(P3\) magnetic symmetry with two uncompensated moments {[}ESI Fig.
10{]}, particularly as analogous structures with different metals may
adopt different ground states. \(P3\) has no inversion centre and is
polar, and so this structure (if it can be engineered in analogues or
through application of stimulus \emph{e.g.} strain) would have
electrical polarisation produced directly by magnetic order (Type II
multiferroicity).\citep{khomskii_classifying_2009} Generally, type II
multiferroics adopt complex magnetic orderings, for example helical
incommensurate states,\citep{kimura_inversion-symmetry_2006} or with
structural triangular lattice arrangements which introduce frustration,
\emph{e.g.} triangular \ce{Sr3NiTa2O9}\citep{liu_two-step_2016} and
\ce{Ba3MnNb2O9}.\citep{lee_magnetic_2014} It would be unusual for a
collinear magnetic ordering to induce this behaviour, and even more so
for a molecular framework. These results illustrate the importance of
magnetic structure determination for understanding the function of
magnetic molecular framework compounds, as also shown by other recent
works on frameworks containing both
Mn(II)\citep{bulledGeometricFrustrationTrillium2022, calderMagneticOrderTwodimensional2023, Walker2017, mansonLongRangeMagneticOrder2001}
and other
metals.\citep{pitcairnLowDimensionalMetalOrganic2023, lopez-cabrellesIsoreticularTwodimensionalMagnetic2018, canadillas-delgadoExperimentalEvidenceCoexistence2020}

In conclusion, we have determined the low temperature structure and
magnetic properties of \ce{[Na(OH2)3]Mn(NCS)3} by developing crystal
growth methodology for this humidity sensitive compound. Neutron
diffraction data permitted the hydrogen atoms to be located and hence
the key role of the hydrogen bonding network in framework structuration.
Refinement of single crystal neutron diffraction data shows that the
compound magnetically orders in \(P \bar{3}'\) magnetic space group,
supported by SHG measurements for which we saw no signal at the
wavelengths we excited, consistent with this centrosymmetric magnetic
structure. The unusual low-symmetry trigonal space group exhibited in
this compound suggest that investigations into the analogues
incorporating metal ions which exhibit magnetic anisotropy (\emph{e.g.}
Co(II), Ni(II), Fe(II)) may lead to the complex spin structure and hence
multiferroic behaviour. Substitutions for metals which form metal
thiocyanate frameworks less sensitive to ambient humidity, \emph{e.g.}
Ni(II), may also lead to more straightforward methods to grow and store
these compounds.

\section*{Conflicts of interest}
No conflicts of interest to declare.

\section*{Acknowledgements}
M.G. acknowledges the ILL Graduate School for provision of a
studentship. M.J.C. acknowledges funding from UKRI (EP/X042782/1) and
the Hobday Bequest School of Chemistry, University of Nottingham. We
acknowledge the ILL for beamtime under proposal numbers 5-41-1167. Raw
data sets from ILL experiments can be accessed via links provided in
references. SHG measurements were performed using equipment funded by
EPSRC strategic equipment grant EP/P001394/1.

\section*{Author Contributions}
M.G. and E.B. synthesised the samples; M.G. grew the crystals; M.G. and
M.J.C. carried out the magnetic measurements and the single crystal
X-ray experiments; M.G., O.F., L.C.D. and M.J.C. carried out the neutron
diffraction experiments and magnetic property analysis; T.B.G. and
A.D.B. carried out the second harmonic generation measurements; M.G.,
L.C.D. and M.J.C. wrote the paper with contributions from all the
authors.


\balance 
\scriptsize{
\bibliography{NaMnNCS3paper.bib} 
\bibliographystyle{rsc} } 
\end{document}